%% file: main.tex
\documentclass[sigconf]{acmart}
\AtBeginDocument{%
  }

\setcopyright{acmlicensed}
\copyrightyear{2018}
\acmYear{2018}
\acmDOI{XXXXXXX.XXXXXXX}
\acmConference[Conference acronym 'XX]{Make sure to enter the correct
  conference title from your rights confirmation email}{June 03--05,
  2018}{Woodstock, NY}
\acmISBN{978-1-4503-XXXX-X/2018/06}

\usepackage{enumitem}
\usepackage{booktabs}
\usepackage{multirow}
\usepackage{tabularx}
\usepackage{makecell}
\usepackage{graphicx}
\usepackage{pifont}

\begin{document}


\title{A Benchmark for Language Models in Real-World System Building}


\author{Weilin Jin}
\authornote{Both authors contributed equally to this research.}
\affiliation{%
  \institution{Peking University}
  \city{Beijing}
  \country{China}
}

\author{Chenyu Zhao}
\authornotemark[1]
\affiliation{%
  \institution{Nankai University}
  \city{Tianjin}
  \country{China}
}

\author{Zeshun Huang}
\affiliation{%
  \institution{Nankai University}
  \city{Tianjin}
  \country{China}
}

\author{Chaoyun Zhang}
\affiliation{%
  \institution{Microsoft}
  \city{Beijing}
  \country{China}
}

\author{Qingwei Lin}
\affiliation{%
  \institution{Microsoft}
  \city{Beijing}
  \country{China}
}

\author{Chetan Bansal}
\affiliation{%
  \institution{Microsoft}
  \city{Redmond}
  \country{USA}
}

\author{Saravan Rajmohan}
\affiliation{%
  \institution{Microsoft}
  \city{Redmond}
  \country{USA}
}

\author{Shenglin Zhang}
\affiliation{%
  \institution{Nankai University}
  \city{Tianjin}
  \country{China}
}

\author{Yongqian Sun}
\affiliation{%
  \institution{Nankai University}
  \city{Tianjin}
  \country{China}
}

\author{Dan Pei}
\affiliation{%
  \institution{Tsinghua University}
  \city{Beijing}
  \country{China}
}

\author{Yifan Wu}
\affiliation{%
  \institution{Peking University}
  \city{Beijing}
  \country{China}
}

\author{Tong Jia}
\affiliation{%
  \institution{Peking University}
  \city{Beijing}
  \country{China}
}

\author{Ying Li}
\affiliation{%
  \institution{Peking University}
  \city{Beijing}
  \country{China}
}

\author{Zhonghai Wu}
\affiliation{%
  \institution{Peking University}
  \city{Beijing}
  \country{China}
}

\author{Minghua Ma}
\authornote{Minghua Ma is the corresponding author. minghuama@microsoft.com}
\affiliation{%
  \institution{Microsoft}
  \city{Redmond}
  \country{USA}
}


\begin{abstract}

\input{sections/0.abstract}
\end{abstract}



\keywords{Cross-ISA Migration, Software Package Build Repair, Benchmark, Model Context Protocol}



\maketitle

\section{INTRODUCTION}
\input{sections/1.introduction}

\section{METHODOLOGY}

\input{sections/2.method}

\section{Experiments}

\input{sections/3.experiments}

\section{DISCUSSION}

\input{sections/4.discussion}


\section{RELATED WORK}

\input{sections/6.related_work}

\section{CONCLUSION}

\input{sections/7.conclusion}


\bibliographystyle{ACM-Reference-Format}
\bibliography{main}






\end{document}

%% file: sections/0.abstract.tex
During migration across instruction set architectures (ISAs), software package build repair is a critical task for ensuring the reliability of software deployment and the stability of modern operating systems. 
While Large Language Models (LLMs) have shown promise in tackling this challenge, prior work has primarily focused on single instruction set architecture (ISA) and homogeneous programming languages. 
To address this limitation, we introduce a new benchmark designed for software package build repair across diverse architectures and languages. 
Comprising 268 real-world software package build failures, the benchmark provides a standardized evaluation pipeline. 
We evaluate six state-of-the-art LLMs on the benchmark, and the results show that cross-ISA software package repair remains difficult and requires further advances. 
By systematically exposing this challenge, the benchmark establishes a foundation for advancing future methods aimed at improving software portability and bridging architectural gaps.
Our benchmark is accessible at \url{https://github.com/zcyyc/Build-bench}.

%% file: sections/1.introduction.tex
\begin{figure*}[t]
  \centering
  \includegraphics[width=0.9\textwidth]{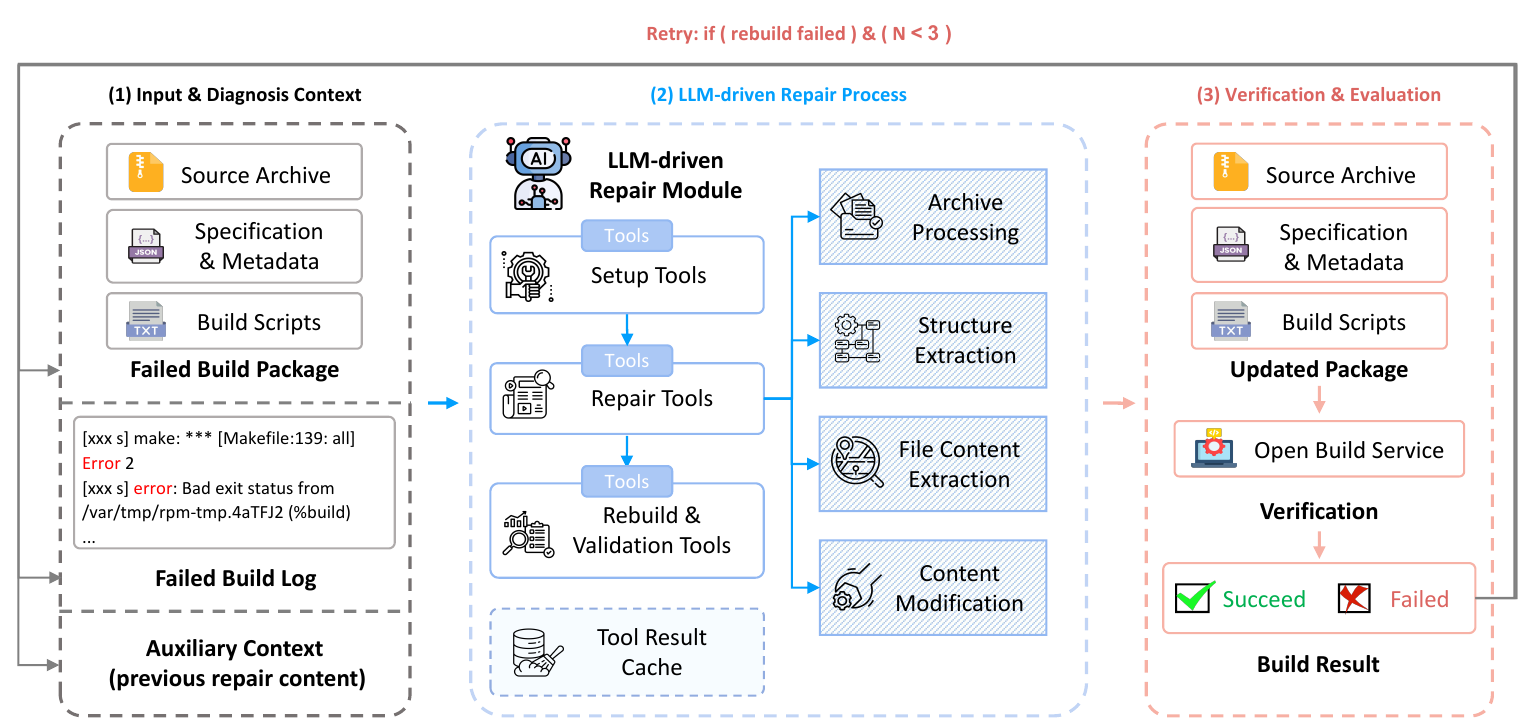}
  \Description{The automatic cross-ISA build repair pipeline of the benchmark.}
  \caption{The automatic cross-ISA build repair pipeline of the benchmark.}
  \label{fig:framework}
\end{figure*}


The emergence of large language models (LLMs) has revolutionized the field of software engineering. By integrating LLMs into modern development workflows, the software ecosystem has evolved toward greater automation, intelligence, and efficiency. Within the field of software engineering, LLMs have achieved impressive results in tasks such as
code generation \cite{DBLP:journals/tosem/DongJJL24}, 
failure prediction \cite{DBLP:conf/issre/LiuM0L0LLCDBRL024}, 
anomaly detection \cite{DBLP:journals/corr/abs-2405-15370}, 
failure diagnosis \cite{DBLP:journals/corr/abs-2407-01710}, 
and program repair \cite{DBLP:conf/icse/BouzeniaDP25}, 
showing great potential to improve how software is developed.

To systematically evaluate the strengths and limitations of large language models in code-related tasks, researchers have proposed a series of benchmarks. For code generation, CoderEval \cite{DBLP:conf/icse/YuSRZZMLLWX24} establishes an independent execution platform for automated evaluation. In root cause analysis, OpenRCA \cite{DBLP:conf/iclr/XuZZHZLPHZ025} is designed to assess the ability of LLMs to analyze software failure causes in real-world environments. In addition, AIOpsLab \cite{DBLP:journals/corr/abs-2501-06706} has been established to evaluate AI agents for enabling autonomous cloud operations. For real-world software engineering tasks, SWE-bench \cite{DBLP:conf/iclr/JimenezYWYPPN24} has been introduced to evaluate the capability of LLMs in resolving practical development issues. However, these benchmarks primarily focus on software engineering problems within homogeneous software and hardware environments, whereas modern software ecosystems often involve components built on heterogeneous instruction set architectures (ISAs). This heterogeneity introduces new challenges for model generalization and adaptability.

Migrating large codebases to new instruction set architectures is of great significance for the evolution of modern software systems. The transitions led by Apple, from PowerPC to x86 and later to ARM \cite{Thornburg01}, together with Amazon’s adoption of Graviton processors \cite{AWSGraviton2023}, exemplify this architectural evolution. Such migrations are mainly driven by three factors: modernizing software that relies on deprecated instruction set architectures, enabling transitions between mobile and desktop platforms, and supporting cloud hyperscalers in moving codebases from x86 to ARM to optimize cloud computing performance. Consequently, ensuring the portability of large-scale open-source software has become an urgent and critical need within the software engineering community.


Despite the growing interest in cross-platform development, repairing software packages across different instruction set architectures remains an extremely challenging task. Such repair involves not only recompilation and adaptation of build systems, but also resolving deep architectural dependencies embedded in source code, libraries, and toolchains. Traditionally, cross-ISA repair has been a complex and labor-intensive process, often requiring package maintainers to manually diagnose build failures, reconfigure dependencies, and modify source code. With the advent of large language models (LLMs), however, the automation of such repair processes has become increasingly feasible.

However, no prior work has evaluated the capability of LLMs in cross-ISA software package build repair. To fill this gap, we propose a new benchmark that enables end-to-end assessment of LLMs in autonomously repairing software packages across heterogeneous instruction set architectures. 
Leveraging standardized orchestration frameworks such as the Model Context Protocol (MCP) \cite{hou2025modelcontextprotocolmcp}, the proposed benchmark takes build-failed packages across different ISAs as input, and employs a variety of external tools, including Structure Extraction, File Content Extraction, and Content Modification tools to analyze and modify source code toward automated repair. During the validation phase, the benchmark utilizes the open platform called Open Build Service (OBS) to perform online package rebuilding, and the rebuilding results are used to determine whether the repair is successful. Furthermore, the benchmark integrates a retry mechanism that triggers additional repair attempts following each failure, with a maximum of three iterations, and terminates once a successful rebuild is obtained. The overall pipeline of the benchmark is illustrated in Figure \ref{fig:framework}.

Based on this framework, we collected 268 real-world software packages from the \texttt{aarch64} and \texttt{x86\_64} architectures, including 163 packages that built successfully on \texttt{x86\_64} but failed when migrated to \texttt{aarch64}, and 105 packages exhibiting the reverse pattern. On these cross-ISA repair tasks, we evaluated six state-of-the-art LLMs which are \textit{GPT-5}, \textit{GPT-5-mini}, \textit{GPT-4o}, \textit{Claude Sonnet 4.5}, \textit{DeepSeek V3}, and \textit{Qwen3-max}. Among them, GPT-5 achieved the highest performance, with a repair success rate of \textbf{63.19\%} in the \texttt{x86\_64→aarch64} direction and \textbf{29.52\%} in the reverse direction. While GPT-5 exhibited strong repair capability, other models performed considerably worse, being limited to resolving only simple dependency or configuration issues. These results highlight that cross-ISA package build repair remains a highly challenging yet promising research direction, with significant potential for further advancement. Overall, the proposed benchmark establishes a solid foundation for evaluating and advancing automated software package repair across heterogeneous architectures.

In summary, our main contributions can be summarized as follows:
\begin{itemize}[leftmargin=1.5em, itemsep=3pt, topsep=2pt]
    \item \textbf{A new benchmark and corpus for cross-architecture build repair.} 
We propose the first benchmark to evaluate the capability of LLMs in software package repair across heterogeneous ISAs. The benchmark includes 268 real-world cases, providing a solid foundation for future research on cross-ISA software repair.
    \item \textbf{An end-to-end evaluation framework with iterative build verification.} 
We provide an integrated evaluation framework that incorporates multiple tools and employs iterative attempts to assess the capability of LLMs in addressing cross-ISA build repair tasks.
    \item \textbf{Comprehensive empirical analysis and quantitative insights.}
We evaluated six state-of-the-art LLMs, and the results demonstrate significant performance disparities across models. These findings expose persistent challenges in multi-file reasoning and architecture adaptation, while establishing a foundation for future research on cross-ISA build repair.    
\end{itemize}

%% file: sections/2.method.tex

\subsection{Overview}

Our benchmark consists of 268 real-world software packages that failed to migrate between the \texttt{aarch64} and \texttt{86\_64} architectures. Based on these cases, we design a complete evaluation pipeline to assess whether LLMs can accurately locate and repair build failures arising from cross-architecture software migration. The entire pipeline is built upon the Model Context Protocol (MCP) and incorporates multiple agent tools responsible for package structure extraction, content extraction, and automated repair. After each repair iteration, the system performs build verification to validate the success of the fix. As illustrated in Figure \ref{fig:framework}, the overall workflow is organized into three main components:
\textit{(1) Input \& Diagnosis Context},
\textit{(2) LLM-driven Repair Module}, and
\textit{(3) Verification \& Evaluation}.

\textbf{Input \& Diagnosis Context}. The input context consists of the source code, configuration files, build scripts, and build failure logs of each software package. 
Moreover, after each failed repair attempt, the input for the next iteration is augmented with the log file generated from the previous attempt. This iterative design enables LLMs to perceive the changes made in each round of repair and progressively refine its strategy, rather than restarting from scratch every time.

\textbf{LLM-driven Repair Module}. We employ the MCP protocol to orchestrate a LLM-driven module for software package build repair. Leveraging the interoperability provided by MCP, the pipeline integrates a diverse set of heterogeneous external tools.
A detailed description of this module is provided in \textbf{Section \ref{subsec:llm-driven-repair-process}}.

\textbf{Verification \& Evaluation}. The packages repaired by LLMs are uploaded to the Open Build Service (OBS) platform for reconstruction. The pipeline determines whether the repair was successful based on the build results returned by OBS. If the build fails and the maximum iteration threshold has not been reached, the pipeline starts a new iteration. At the beginning of each new iteration, updated build logs replace the original package information, while the repair logs from the previous iteration are appended as additional context. This iterative workflow enables a thorough evaluation of the model’s end-to-end capability in software package migration and repair.

\subsection{Benchmark Construction}
\label{subsec:benchmark-construction}

For benchmark construction, we collect software packages from the \textbf{Open Build Service} (OBS) repositories, where a total of 17,001 packages successfully build on \texttt{x86\_64} and 16,892 packages successfully build on \texttt{aarch64}. Given the large number of available packages, 6,000 packages from each architecture are randomly sampled as the source sets. Based on these sets, packages that fail to build on the opposite architecture are identified as candidate samples. To ensure reproducibility, each failed package is rebuilt on OBS, and only those which consistently fail during the repeated build process are retained. As a result, the final benchmark comprises 268 cross-ISA build failures, including 163 packages that build successfully on \texttt{x86\_64} but fail on \texttt{aarch64}, and 105 packages that succeed on \texttt{aarch64} but fail on \texttt{x86\_64}.

\begin{figure}[t]
  \centering
  \setlength{\abovecaptionskip}{3pt}
  \setlength{\belowcaptionskip}{-10pt}
  \includegraphics[width=\columnwidth]{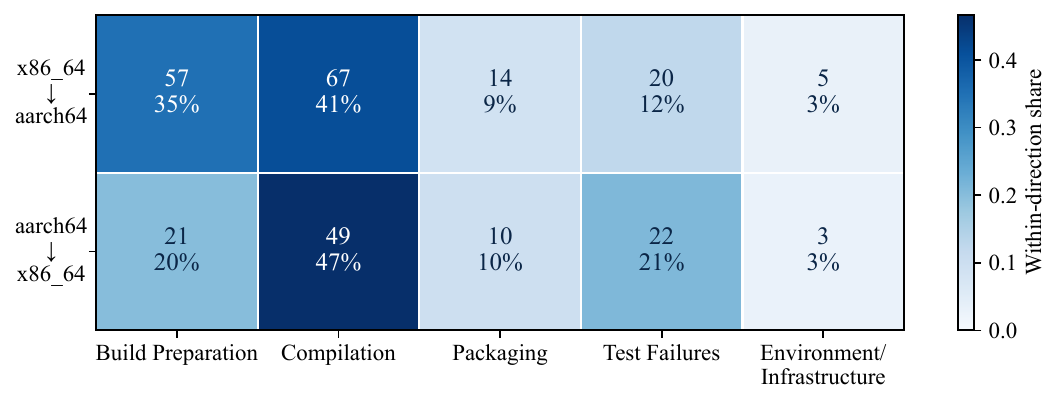}
  \caption{Error category distribution across cross-ISA build failures. 
  The figure illustrates the proportion of five error types across both migration directions (\texttt{x86\_64→aarch64} and \texttt{aarch64→x86\_64}).}
  \label{fig:error_distribution}
\end{figure}

To better illustrate the composition of the benchmark,we employ GPT-5-mini to analyze the build logs of failed packages and categorize them into five specific error types.
These categories, summarized in Figure~\ref{fig:error_distribution}, are described as follows:

\begin{itemize}[leftmargin=1.5em, itemsep=3pt, topsep=2pt]
    \item \textbf{Build Preparation Errors.} Typically occur due to missing dependencies, misconfigured toolchains, or invalid compiler flags.
    
    \item \textbf{Compilation Errors.} Arise from language incompatibilities or mismatches between build systems.
    
    \item \textbf{Packaging Errors.} Result from incomplete or incorrectly specified build artifacts. 
    
    \item \textbf{Test Failures.} Occur when the software produces unexpected behavior or fails to pass verification tests.
    
    \item \textbf{Environment or Infrastructure Errors.} Caused by external factors such as virtual machine shutdowns or resource interruptions.
\end{itemize}

These error patterns can be viewed from multiple perspectives. From the build process perspective, they span the entire software construction pipeline — including environment setup, dependency resolution, compilation, packaging, and runtime verification; From the build content perspective, they involve highly heterogeneous information sources such as configuration scripts (e.g., \texttt{CMakeLists}, \texttt{Makefile}), build specifications, source code files, dependency manifests, and testing environments. Together, these dimensions reveal that cross-ISA migration is not a simple recompilation task but rather a multi-stage reasoning problem that requires LLMs to analyze complex build contexts, understand dependency interactions, and adaptively integrate knowledge across both code and environment domains. This intrinsic complexity underscores the necessity of comprehensive reasoning and appropriate tool use in automated build repair.

\subsection{LLM-driven repair process}
\label{subsec:llm-driven-repair-process}

To evaluate the end-to-end software package build repair capability of LLMs, the benchmark framework connects to MCP servers. Unlike static pipelines that follow fixed execution sequences, the MCP architecture is context-aware: different tools share a unified context, and the system can dynamically orchestrate tool invocations, allowing LLMs to autonomously determine the order of operations. This design transforms the repair workflow into an LLM-driven reasoning process, allowing the model to dynamically coordinate and utilize external tools for analysis and repair. Consequently, the framework enables direct evaluation whether an LLM can effectively handle software package migration and repair tasks.

\subsubsection{MCP Tools}

MCP tools integrated into the benchmark framework are divided into three main categories: \textbf{Setup Tools}, \textbf{Repair Tools}, and \textbf{Rebuild \& Validation Tools}. Setup Tools are responsible for environment initialization, retrieving target packages files, and maintaining the tool result cache; Repair Tools perform file inspection and modification tasks; Rebuild \& Validation Tools handle package repackaging, uploading to the build platform, and verifying the build results.

Among these, Repair Tools play a pivotal role in the benchmark, consisting of four key modules that support code inspection and modification throughout the repair workflow. 
Each of these modules is designed to address a specific aspect of the repair process, as described below:

\begin{itemize}[leftmargin=1.5em, itemsep=3pt, topsep=2pt]
    \item \textbf{Archive Processing Tool} manages the extraction and compression of source archives. To prevent unintended modifications to the original source files, it allows the LLM to extract the code into a temporary directory, apply the required changes, and then repackage the files into a format compatible with the build service.

    \item \textbf{Structure Extraction Tool} uses a recursive approach to scan the hierarchical structure of the source files. During this process, non-essential documentation and configuration files (e.g., README.md) are excluded, and a JSON-formatted structural representation is generated. It also identifies class and function definitions within the code and determines their boundaries.

    \item \textbf{File Content Extraction Tool} retrieves the full content of target files to ensure that the LLM operates on complete code rather than truncated fragments.

    \item \textbf{Content Modification Tool} supports two modes of file editing: Full File Generation, which regenerates the entire file while preserving its structural layout, and Patch Generation, which applies line-level diff-style edits such as additions, deletions, and modifications.
     
\end{itemize}

In addition, Rebuild \& Validation Tools are responsible for uploading the rebuilt software packages to the OBS platform and verifying the corresponding build results. Once the build process is triggered on the platform, the tools periodically poll for build results. Instead of returning a simple binary success or failure signal, the system categorizes each result into one of four explicit states: \textit{succeeded} (the package is successfully rebuilt), \textit{broken} (missing or malformed build descriptors such as corrupted specification files), \textit{unresolvable} (missing or undefined dependencies), and \textit{failed} (general build failure). This detailed feedback design provides stable and verifiable diagnostic signals for each iteration, facilitating more effective resolution of build issues throughout the repair process. Ultimately, Rebuild \& Validation Tools ensure that package repair is not only syntactically correct but also verifiable in the target environment.

\subsubsection{Tool Result Cache}

To reduce token consumption and improve inference efficiency, this component is utilized in every repair attempt. During the reasoning process, some LLMs tend to invoke the same tool multiple times, even when repeated calls yield nearly identical results. For tools such as Structure Extraction, which produce highly consistent outputs across repeated invocations, \textbf{a tool result cache} is crucial to avoid unnecessary recomputation. The caching mechanism prevents redundant computations, accelerates execution, and ensures more efficient use of the context window.

It is worth noting that, to ensure each iteration is executed with a clean and consistent tool state, a dedicated reset package cache tool is invoked at the beginning of every new iteration. This tool clears the cache generated in the previous round and resets the workspace environment, enabling the LLM to regenerate tool outputs from scratch under a controlled and reliable setup.

\subsubsection{Iterative Reasoning and Verification Loop}

The repair process in the benchmark is not a single-turn interaction but an iterative reasoning loop that incorporates build feedback into subsequent repair attempts. Each repair iteration is performed incrementally on the current repaired state, rather than restarting from the original package, thereby enabling cumulative and stateful repair across iterations. After each round of modification, the updated package is uploaded to the Open Build Service (OBS) platform for reconstruction. This provides a reproducible and verifiable environment for cross-ISA repair tasks, where accurate build results and detailed build logs are generated for every iteration. Based on the results returned by the platform, the system determines whether another iteration of repair should be initiated.

If the build fails in the current iteration and the maximum iteration limit ($N_{\max}$) has not been reached, the framework reinitiates the loop by updating three key inputs:
(1) the raw failed build package,
(2) the latest build log, and
(3) the auxiliary context, which records previous repair actions and defines the scope of prior modifications.
These inputs are integrated into the reasoning context. During this process, the framework replaces the previous build log with the latest one, ensuring that failure analysis is always based on the latest build evidence while avoiding interference from redundant or outdated information.

Within this iterative loop, LLM autonomously repairs build failures caused by cross-architecture migration and dynamically determines when and how to invoke the appropriate tools to modify build scripts or source code. This iterative design preserves the history of prior attempts, providing the LLM with richer contextual information and enabling progressive reasoning under real feedback. As a result, the benchmark evaluates not only the model’s static repair capability, but also its ability to adapt, reflect, and converge toward a successful cross-architecture build.

%% file: sections/3.experiments.tex
In the context of cross-ISA software package build repair, we conducted a comprehensive evaluation of six representative large language models, including \textit{GPT-5}, \textit{GPT-5-mini}, \textit{GPT-4o}, \textit{Claude Sonnet 4.5}, \textit{Qwen3-max}, and \textit{DeepSeek-V3}, to demonstrate the effectiveness of the proposed benchmark. Specifically, experiments are designed to address the following research questions:

\textbf{RQ1}: How well do current LLMs perform in repairing cross-architecture build failures?  

\textbf{RQ2}: Does iterative feedback from build logs improve LLM repair performance over single-turn evaluation?  


To ensure a fair comparison, all models are evaluated under identical experimental settings, including the same package corpus, prompt templates, maximum number of tool invocations ($T_{max}=20$), and iteration limit ($N_{max}=3$). The temperature is fixed at 0 for all models throughout the evaluation. All models are evaluated without enabling any thinking or extended reasoning modes. Consequently, token usage counts only user-visible input and output tokens.

\subsection{Evaluation Metrics}

We evaluate the repair performance of LLMs from two perspectives: effectiveness and efficiency. The evaluation includes the following metrics:

\begin{itemize}[leftmargin=1.5em, itemsep=3pt, topsep=2pt]
    \item \textbf{Build Success Rate (\%)}. The percentage of successfully built packages among all packages included in the benchmark for the given architecture.
    \item \textbf{Average Repair Time (min)}. The average time required for the model to repair a package or reach the maximum number of iterations.
    \item \textbf{Average Token Consumption (K)}. The average number of tokens (input and output tokens) consumed by the model during the repair process of each package.
\end{itemize}



\begin{table*}[t]
  \centering
  \caption{Repair performance of LLMs on cross-ISA migration failures in both directions. 
    Here, \textit{Time} indicates the average repair time (minutes), 
    \textit{Tokens} refers to the average token usage, 
    and \textit{Avg. Iter} denotes the average successful repair iteration conditioned on final success.}
  \label{tab:cross-isa-performance}
  \setlength{\tabcolsep}{3pt}
  \renewcommand{\arraystretch}{1.1}

  \begin{tabular}{l|ccccc|ccccc}
    \toprule
    & \multicolumn{5}{c|}{\textbf{x86\_64 $\rightarrow$ aarch64 (163 Packages)}} &
      \multicolumn{5}{c}{\textbf{aarch64 $\rightarrow$ x86\_64 (105 Packages)}} \\
    \cmidrule(lr){2-6} \cmidrule(lr){7-11}
    \makecell[c]{\textbf{Model}\\~} &
      \textbf{\#Succ.} & \textbf{Succ. (\%)} & \textbf{Time (min)} & \textbf{Tokens (K)} & \textbf{Avg. Iter} &
      \textbf{\#Succ.} & \textbf{Succ. (\%)} & \textbf{Time (min)} & \textbf{Tokens (K)} & \textbf{Avg. Iter} \\
    \midrule
    GPT-5             & \textbf{103} & \textbf{63.19} & 31.18 & 1830.91 & 1.65 &
                        \textbf{31}  & \textbf{29.52} & 18.55 & 1518.66 & 2.06 \\
    GPT-5-mini        & 47  & 28.83 & 13.80 & 1683.95 & 1.77 &
                        28  & 26.67 & 14.37 & 1894.60 & 1.86 \\
    GPT-4o            & 22  & 13.50 & 5.93  & 541.66  & 1.68 &
                        13  & 12.38 & 5.82  & 614.12  & 2.23 \\
    Claude Sonnet 4.5 & 16  & 9.82  & 6.27  & 328.76  & 1.00 &
                        6   & 5.71  & 4.52  & 332.99  & 1.33 \\
    DeepSeek V3       & 13  & 7.98  & 11.37 & 235.53  & 1.77 &
                        4   & 3.81  & 19.27 & 445.03  & 2.50 \\
    Qwen3-max         & 28  & 17.18 & 64.69 & 505.39  & 2.07 &
                        6   & 5.71  & 52.44 & 714.08  & 2.67 \\
    \bottomrule
  \end{tabular}
\end{table*}

\subsection{RQ1: Performance of LLMs on the Benchmark}

To evaluate the effectiveness of LLMs in cross-ISA software package build repair, we conducted experiments on six representative models, as summarized in Table~\ref{tab:cross-isa-performance}. The results indicate that current models still exhibit significant gaps in repair accuracy, task completion time, and token efficiency, highlighting the effectiveness and discriminative power of the proposed benchmark.
In the following sections, we present a detailed analysis of the results from two perspectives: \textbf{Cross-ISA Repair Performance} and \textbf{Time and Token Efficiency}, which together provide a comprehensive view of model capability and efficiency across heterogeneous architectures.

\subsubsection{Cross-ISA Repair Performance} In terms of Cross-ISA Repair Accuracy, GPT-5 achieved the best performance among all models. For the migration direction from \texttt{x86\_64} to \texttt{aarch64}, it successfully repaired 103 out of 163 packages, reaching an accuracy of 63.19\%, which is significantly higher than that of GPT-5-mini (28.83\%) and Qwen3-max (17.18\%). With the exception of GPT-5, all other models failed to reach a high level of accuracy, with Claude Sonnet 4.5 and DeepSeek V3 performing particularly poorly—both below 10\%. This suggests that these models are largely incapable of handling complex package repair tasks in cross-architecture settings. 

In the opposite direction (\texttt{aarch64} $\rightarrow$ \texttt{x86\_64}), GPT-5 repaired 31 out of 105 packages, achieving 29.52\% accuracy, which is comparable to GPT-5-mini (26.67\%) and GPT-4o (12.38\%). The remaining models performed considerably worse, with success rates dropping below 10\%, indicating their limited capability in handling cross-ISA build repair.

Overall, the results indicate that LLMs exhibit stronger repair capability in the \texttt{x86\_64} $\rightarrow$ \texttt{aarch64} migration direction, which could be attributed to recent industrial and academic efforts have predominantly targeted this transition, allowing models to accumulate more implicit knowledge related to ARM-based environments. Conversely, performance on the reverse migration path remains weaker, as this direction has received comparatively less attention. These findings further reveal that the difficulty of repair varies between migration directions, underscoring the asymmetric nature of cross-ISA adaptation.

\subsubsection{Time and Token Efficiency} To assess the resource cost, we measure the average time and number of tokens each model consumes to repair a package. The experimental results reveal substantial variations in both repair time and token consumption across models, reflecting significant differences in their inference strategies.

Overall, most models achieve satisfactory results in both time and token usage. Although GPT-5 and GPT-5-mini consume more time and tokens, they achieve significantly higher build repair accuracy, demonstrating that the additional computational cost translates into improved repair effectiveness. In contrast, GPT-4o and Claude Sonnet 4.5 show faster inference and lower token consumption, while exhibiting lower repair accuracy, making them more suitable for time-sensitive repair scenarios. Despite its minimal token usage, DeepSeek V3 shows suboptimal performance in both accuracy and efficiency, which may be attributed to differences in model architecture and pretraining data compared with newer LLMs.

However, Qwen3-max represents a notable exception among the evaluated models. It achieves the longest average runtime, taking around 60 minutes to repair each of the 268 packages on average. Log analysis reveals that most of this overhead originates from repeatedly invoking the \textit{Check Build Result tool} for verification, where excessive validation significantly extends the repair cycle without improving the final success rate.

\subsection{RQ2: Impact of Iterative Feedback}

\begin{table*}[t]
  \centering
  \caption{Iteration-wise improvement in repair success rates on the benchmark. 
  $\Delta$(2–1) and $\Delta$(3–2) indicate incremental improvements between adjacent iterations, 
  and $\Delta$(3–1) denotes cumulative improvement.}
  \label{tab:iterative-feedback}
  \setlength{\tabcolsep}{5pt}
  \renewcommand{\arraystretch}{1.1}

  \begin{tabular}{l|cccccc|cccccc}
    \toprule
    & \multicolumn{6}{c|}{\textbf{x86\_64 $\rightarrow$ aarch64 (163 Packages)}} &
      \multicolumn{6}{c}{\textbf{aarch64 $\rightarrow$ x86\_64 (105 Packages)}} \\
    \cmidrule(lr){2-7} \cmidrule(lr){8-13}
    \makecell[c]{\textbf{Model}\\~} &
      \textbf{Iter-1 (\%)} & \textbf{Iter-2} & \textbf{Iter-3} & \textbf{$\Delta$(2–1)} & \textbf{$\Delta$(3–2)} & \textbf{$\Delta$(3–1)} &
      \textbf{Iter-1 (\%)} & \textbf{Iter-2} & \textbf{Iter-3} & \textbf{$\Delta$(2–1)} & \textbf{$\Delta$(3–2)} & \textbf{$\Delta$(3–1)}  \\
    \midrule
    GPT-5             & 36.81 & 48.47 & 63.19 & $\uparrow$\textbf{11.66} & $\uparrow$\textbf{14.72} & $\uparrow$\textbf{26.38} &
                        11.43 & 16.19 & 29.52 & $\uparrow$4.76 & $\uparrow$\textbf{13.33} & $\uparrow$\textbf{18.10} \\
    GPT-5-mini        & 15.34 & 20.25 & 28.83 & $\uparrow$4.91 & $\uparrow$8.58 & $\uparrow$13.50 &
                        12.38 & 18.10 & 26.67 & $\uparrow$5.72 & $\uparrow$8.57 & $\uparrow$14.29 \\
    GPT-4o            & 4.91 & 12.88 & 13.50 & $\uparrow$7.97 & $\uparrow$0.62 & $\uparrow$8.59 &
                        0.95 & 8.57 & 12.38 & $\uparrow$\textbf{7.62} & $\uparrow$3.81 & $\uparrow$11.43 \\
    Claude Sonnet 4.5 & 9.82 & 9.82 & 9.82 & $\uparrow$0 & $\uparrow$0 & $\uparrow$0 &
                        4.76 & 4.76 & 5.71 & $\uparrow$0 & $\uparrow$0.95 & $\uparrow$0.95 \\
    DeepSeek V3       & 3.68 & 6.13 & 7.98 & $\uparrow$2.45 & $\uparrow$1.85 & $\uparrow$4.29 &
                        0.95 & 0.95 & 3.81 & $\uparrow$0 & $\uparrow$2.86 & $\uparrow$2.86 \\
    Qwen3-max   & 6.13 & 9.82 & 17.18 & $\uparrow$3.69 & $\uparrow$7.36 & $\uparrow$11.04 &
                        0.95 & 0.95 & 5.71 & $\uparrow$0 & $\uparrow$4.76 & $\uparrow$4.76 \\
    \bottomrule
  \end{tabular}
\end{table*}

Table \ref{tab:iterative-feedback} summarizes the impact of iterative feedback on the final repair success rate.
In each iteration, the build logs are updated and the previous repair information is incorporated into the next round of reasoning. The results show that repair accuracy improves with the number of iterations.

Overall, iterative feedback substantially improves performance across most models. In the \texttt{x86\_64} → \texttt{aarch64} direction, GPT-5 and GPT-5-mini improve their repair success rates by 26.38\% and 13.50\% after three iterations. In the reverse direction (\texttt{aarch64} → \texttt{x86\_64}), the improvements are 18.10\% and 14.29\%. These results clearly demonstrate the effectiveness of iterative refinement.
GPT-4o and Qwen3-max also exhibit notable improvements, indicating that iterative feedback remains beneficial even when a model’s reasoning capability is relatively limited.
In contrast, DeepSeek V3 and Claude Sonnet 4.5 show only marginal improvements. Notably, Claude Sonnet 4.5 achieves no increase for the \texttt{x86\_64} → \texttt{aarch64} task.
Log analysis reveals that Claude Sonnet 4.5 typically modifies the target files only during the first iteration and, in subsequent iterations, merely rechecks and validates the same files without performing further edits. This behavior prevents the model from effectively leveraging newly generated repair logs for targeted corrections, directly limiting its iterative repair performance.

When comparing the improvement between consecutive iterations, noticeable differences emerge across models.
For GPT-5, GPT-5-mini, and Qwen3-max, the improvement achieved in the third iteration exceeds that of the second, indicating that additional iterations continue to enhance repair performance. This finding implies that multi-round iterative feedback enables advanced models to better exploit their reasoning ability and tackle more complex repair cases.
In contrast, GPT-4o, Claude Sonnet 4.5 and DeepSeek V3 exhibit reduced improvement from the second to the third iteration. Due to limitations such as weaker reasoning ability and smaller context windows, these models perform most of their effective repairs within the first two iterations, with little or no improvement observed in the third.

%% file: sections/4.discussion.tex
To evaluate the effectiveness of the proposed benchmark, we analyze whether it accurately captures the key challenges involved in cross-ISA build repair from two perspectives: Representativeness and Extensibility.

\subsection{Representativeness}

The benchmark is constructed from real-world OBS build failures, covering a broad range of migration-related issues across toolchains, build scripts, and packaging configurations.
As described in \textbf{Section \ref{subsec:benchmark-construction}}, selected packages include errors that occur at various stages of the build process, allowing a comprehensive evaluation of LLM performance across different phases of software reconstruction.
Moreover, the observed variation in model performance demonstrates that the benchmark captures realistic cross-ISA complexity and effectively exposes differences in reasoning ability, adaptability, and robustness among LLMs.


\subsection{Generality and Extensibility}


Beyond its representativeness, the proposed benchmark demonstrates strong generality and extensibility from multiple perspectives as follows:

\begin{itemize}[leftmargin=1.5em, itemsep=3pt, topsep=2pt]
    \item \textbf{Error coverage}. The benchmark encompasses five major categories (\textit{ Build Preparation Errors, Compilation Errors, Packaging Errors, Test Failures, and Environment or Infrastructure Errors}) that are not specific to any single architecture but commonly occur in real world build repair scenarios.
    \item \textbf{Input generality}. The benchmark adopts a standardized format that takes software package source files and corresponding build logs as input, making it applicable to a wide range of software ecosystems and build systems.
    \item \textbf{Model integration}. The benchmark leverages the Model Context Protocol (MCP) to decouple LLMs from external agent tools, allowing new models to be seamlessly incorporated into the pipeline without additional adaptation efforts.
    \item \textbf{Result validation}. The benchmark relies on the Open Build Service (OBS), an open-source distributed build and release platform that currently supports not only \texttt{x86\_64} and \texttt{aarch64} but also a wide range of other architectures.
\end{itemize}

Collectively, these design principles ensure that our benchmark serves not only as a reliable dataset for model evaluation but also as a reproducible and extensible framework for advancing research in automated software package build repair across heterogeneous computing environments.

%% file: sections/6.related_work.tex
To provide a clear research context, we first outline the taxonomy of LLM-based code-related tasks, followed by a review of existing program repair benchmarks.

\subsection{LLMs for Code-related Tasks}
\label{subsec:llm-for-code-ralated-task}

Following the taxonomy proposed by Anand et al. \cite{DBLP:journals/corr/abs-2411-07586}, LLM-based code tasks can be broadly divided into two categories: \textbf{Automated Program Repair} and \textbf{Code Generation}.
Recent research highlights that the integration of LLMs into software engineering has significantly advanced code generation, anomaly detection, incident triage, and root cause analysis \cite{li2022constructing, zhong2023survey, zhang2024failure, zhang2024large, chen2024aiopslab, liu2025opseval, mu2025gui, zhao2025can, zhao2025triage}. 

Recently, Large Language Models (LLMs) have advanced Automated Program Repair (APR) from traditional template-based methods (e.g., TBar \cite{DBLP:conf/issta/LiuK0B19}) toward end-to-end, data-driven fixing. Xia et al. \cite{DBLP:conf/icse/XiaWZ23} demonstrated that LLMs outperform heuristic approaches in patch accuracy and compilation rates across diverse languages. Building on this, ThinkRepair \cite{DBLP:conf/issta/Yin00LZ024} introduced a self-directed framework using chain-of-thought and iterative feedback, significantly improving performance on benchmarks like Defects4J \cite{DBLP:conf/issta/JustJE14}.

Meanwhile, code generation research has shifted from simple synthesis to collaborative, tool-augmented paradigms. Recent works like Self-Collaboration \cite{DBLP:journals/tosem/DongJJL24} employ multi-agent roles (analyst, coder, tester) for complex task decomposition, while CodeAgent \cite{DBLP:conf/acl/ZhangLLSJ24} integrates repository-level tools to boost real-world performance. Furthermore, Wu et al. \cite{DBLP:conf/icse/WuWLTYY0L25} validated that in-context learning can effectively generate semantically consistent commit messages. Together, these studies highlight a move toward multi-turn, tool-integrated software development.

Beyond repair and generation, recent studies leverage LLMs for failure diagnosis and root cause analysis (RCA). While traditional approaches \cite{DBLP:conf/kbse/SunSMMXZP24, DBLP:conf/kdd/XieZGZMNYXSLP24, DBLP:conf/issre/LiML0LLCDBRL024, zhao2023robust} excel in fault localization and incident management, LLMs offer superior contextual reasoning and generalization across diverse failure scenarios. For instance, Zhang et al. \cite{DBLP:conf/sigsoft/ZhangGBWM0R24} utilized GPT-4 and in-context learning to automate RCA without fine-tuning. These advancements illustrate the synergy between LLMs and AIOps for intelligent system recovery, inspiring our benchmark's focus on cross-ISA build failure repair to address architectural incompatibilities.

\subsection{Benchmarks for Automated Program Repair}
\label{subsec:benchmarks-for-APR}

We compare representative benchmarks in Automated Program Repair (APR). Early datasets like Defects4J \cite{DBLP:conf/issta/JustJE14} established the foundation for reproducible research using real-world Java bugs. Building on this, GitBug-Java \cite{DBLP:conf/msr/SilvaSM24} curated 199 recent bugs while emphasizing long-term executability via preserved build environments.

More recently, large-scale benchmarks have shifted toward complex, real-world scenarios. SWE-bench \cite{DBLP:conf/iclr/JimenezYWYPPN24} leverages GitHub issues to evaluate multi-file Python repair, while its successor, SWE-bench-Live \cite{DBLP:journals/corr/abs-2505-23419}, introduces Dockerized environments and continuous updates. Additionally, FeedbackEval \cite{DBLP:journals/corr/abs-2504-06939} assesses LLMs’ ability to iteratively refine code based on structured feedback.

Despite these advancements from static datasets to dynamic, feedback-aware evaluations, a gap remains: none of these benchmarks address cross-ISA build repair. While existing works focus on source-level bugs, our benchmark evaluates LLMs’ autonomy in diagnosing and repairing build failures caused by migration between heterogeneous architectures.

%% file: sections/7.conclusion.tex
In this paper, we proposed the first benchmark for evaluating large language models (LLMs) in cross-ISA software package build repair. Built upon the Open Build Service (OBS) platform and Model Context Protocol (MCP) frameworks, it provides an end-to-end evaluation pipeline covering build analysis, automated repair, and iterative validation.
Experiments on six state-of-the-art LLMs show that while current models can fix simple dependency and configuration issues, they struggle with complex, architecture-specific repairs and long-context reasoning. The benchmark highlights these challenges and establishes a solid foundation for future research on architecture-aware and reasoning-efficient LLMs, advancing the automation of cross-ISA software repair and portability.